\documentclass[a4paper,fleqn]{cas-dc}

\usepackage[sort&compress,numbers]{natbib}
\newcommand{\wymiarsmall}{0.4\textwidth}

\begin{document}
\let\WriteBookmarks\relax
\def\floatpagepagefraction{1}
\def\textpagefraction{.001}

\shortauthors{A. Cichy, K. J. Kapcia, A. Ptok}

\title [mode = title]{Spin-polarized superconducting phase in semiconducting system with next-nearest-neighbor hopping on the honeycomb lattice}

\shorttitle{Spin-polarized superconducting phase in semiconducting system with NNN hopping on the honeycomb lattice}

\author[1,2]{Agnieszka Cichy}[%
                    orcid=0000-0001-5835-9807]
\ead{agnieszkakujawa2311@gmail.pl}
\credit{Conceptualization, Methodology, Software, Validation, Formal analysis, Investigation, Resources, Data curation, Writing - Review \& Editing}

\author[1,3]{Konrad Jerzy Kapcia}[%
                    orcid=0000-0001-8842-1886]
\ead{konrad.kapcia@amu.edu.pl}
\credit{Investigation, Writing - Original draft preparation, Writing - Review \& Editing, Visualization, Supervision, Project administration, Funding acquisition}

\author[4]{Andrzej Ptok}[%
                    orcid=0000-0002-5566-2656]
\ead{aptok@mmj.pl}
\credit{Methodology, Investigation, Writing - Original draft preparation, Writing - Review \& Editing, Funding acquisition}

\address[1]{Institute of Spintronics and Quantum Information, Faculty of Physics, Adam Mickiewicz University in Pozna\'n, ul. Uniwersytetu Pozna\'{n}skiego 2, 61614 Pozna\'{n}, Poland}
\address[2]{Institut f\"{u}r Physik, Johannes Gutenberg-Universit\"{a}t Mainz, Staudingerweg 9, 55099 Mainz, Germany}
\address[3]{{Center for Free-Electron Laser Science CFEL, Deutsches Elektronen-Synchrotron DESY,} {Notkestr. 85, 22607, Hamburg, Germany}}
\address[4]{Institute of Nuclear Physics, Polish Academy of Sciences, W. E. Radzikowskiego 152, 31-342 Krak\'{o}w, Poland}

\begin{abstract}
The particles in the honeycomb lattice with on-site $s$-wave pairing exhibit many interesting behaviors, which can be described in the framework of the Hubbard model.
Among others, at the half-filling, some critical value $|U_c|$ of pairing interaction $U<0$ exists that, for $U<U_c$, the superconducting phase becomes unstable.
Introduction of the nonzero hopping $t'$ between next-nearest-neighbor sites strongly modifies the physical properties of the system.
Here, we discuss the behavior of the system for $t' \neq 0$ (at the ground state), where the hopping between next-nearest neighbors leads to change of the order of phase transition between superconducting and normal phases from discontinuous to continuous one in the external magnetic field $h$. 
We show that this behavior is strongly dependent on $t'$ and associated with the Dirac cones in the non-interacting band structure of the system.
For non-zero magnetic field and for some range of $t'$, 
a spin-polarized superconducting phase occurs in the ground state phase diagram (only at the half-filling and for $h\neq 0 $).
\begin{itemize}
\item[$\ $]{\large{\textsc{Highligts:}}}
\item Ground state of the Hubbard model in the external magnetic field is investigated.
\item Effects of the next-nearest-neighbor hopping are investigated on the honeycomb lattice.
\item Two different superconducting phases are found to be stable.
\item The spin-polarized superconducting (so-called Sarma) phase  occurs in the field.
\item The ground state phase diagrams of the model are determined.
\end{itemize}
\end{abstract}

%
%
%
%
%
%
%
\maketitle
%
%
%

\section{Introduction}

The honeycomb lattice, containing two atoms in a primitive unit cell, is characterized by several properties. 
For instance, the electronic structure is formed by two bands, which allows to realize the Dirac cones at the K-points of the Brillouin zone.
The most famous and the simplest example of realization of such a lattice in the nature is graphene~\cite{castroneto.guinea.09}.
Observation of the Dirac physics in that simple lattice of carbon atoms attracted a huge attention not only in a context of fundamental studies, but also in potential applications.
From experimental point of view, the graphene-like lattice exhibits several interesting features.
For example, the following phenomena are worth to mention: a realization of the edge-dependent electronic edge mode in the nanoribbons geometry~\cite{ryu.hatsugai.02,wakabayashi.takane.07,yao.yang.09}, the quantum spin Hall effect (originally formulated for graphene)~\cite{kane.mele.05}, or experimentally observed the quantum Hall
effect~\cite{zhang.tan.05}.
Such unique phenomena open various possibilities of the graphene applications in, e.g., spinotronics~\cite{avsar.ochoa.20} or valleytronics ~\cite{bussolotti.kawai.18}.

Quite recently, a relatively simply way of the graphene multilayer structures manipulation was used for studies of the twisted bilayer graphene lattice~\cite{bistritzer.macdonald.11}. 
It is characterized by the Mori\'e pattern and magic angles~\cite{he.zhou.21}.
In such a case, arbitrary changing of the angle between the layers allows to observe different phenomena like, e.g., unconventional superconductivity~\cite{cao.fatemi.18,lu.setpanov.19,yankowitz.shaowen.19,arora.polski.20,codecido.wang.19,saito.ge.20}, an insulating phase~\cite{codecido.wang.19,saito.ge.20,stepanov.das.20,cao.fatemi.18b}, topological edge states~\cite{zhang.macdonald.13,vaezi.liang.13,ju.shi.15,brown.walet.18}, or the fractional quantum Hall effect~\cite{hunt.sanchezyamagishi.13}.

Also a single honeycomb layer exhibits extraordinary properties in a context of the superconducting states.
For example, the fermionic particles in the atomic Fermi gas on the honeycomb lattice undergo a crossover from the Bardeen--Cooper--Schrieffer (BCS) state to the Bose--Einstein condensation (BEC) of diatomic molecules, shorter named as the {\it BCS--BEC crossover}~\cite{zhao.paramekanti.06}.
Additionally, in the presence of the external magnetic field, the Fulde--Ferrell--Larkin--Ovchinnikov (FFLO) phase can occur~\cite{cichy.ptok.18}.

{\it Motivation.} One of parameters which strongly affects the electronic band structure of the honeycomb lattice is the hopping integral between next-nearest-neighbor (NNN) sites~\cite{cichy.kapcia.22}. 
In the absence of the hopping between NNN sites, the behavior of fermions on the honeycomb lattice at half-filling is characterized by a critical pairing interaction strength $|U_{c}|$, below which the system is semiconducting.
Let us underline that, in this case, the semiconducting (sometimes refereed as semimetallic) behavior is related to: (i) vanishing density of states at the Fermi level and (ii) two bands (the conduction and valence bands), which touch each other at some points in momentum space (i.e., the Fermi surface shrinks to the Dirac points). 
However, for the pairing interaction above $|U_{c}|$, the ground states becomes superconducting.
In this work, we investigate the phase transitions between semiconducting and superconducting phases as well as an influence of the NNN hopping on them.

The paper is organized as follows.
First, the model used in the current study is formulated in Sec.~\ref{sec.model}.
Next, in Sec.~\ref{sec.res} we present and discuss obtained numerical results in the ground state.
Finally, the main conclusions summarize the work in Sec.~\ref{sec.sum}.

\section{Model and approximation}
\label{sec.model}

\begin{figure}[b]
    \centering
    \includegraphics[width=\wymiarsmall]{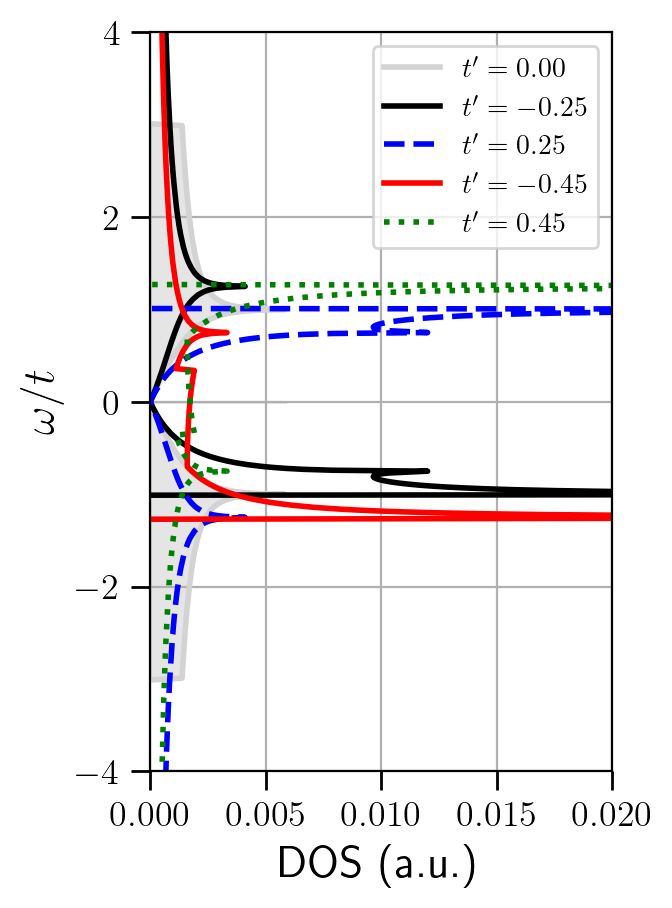}
    \caption{\label{fig.dos}%
        The non-interacting electronic density of states for different values of the hopping between next-nearest-neighbor sites $t'$ (as labeled).
        $\omega =0$ corresponds to the Fermi level at the half-filling.}
\end{figure}

In this work, we investigate the fermionic particles on the honeycomb lattice in the frame of the Hubbard model:
\begin{eqnarray}\label{eq.initialham}
    \nonumber \hat{H} = \sum_{ij\sigma} \left[ -t_{ij} - \left( \mu + \sigma h \right) \delta_{ij} \right] \hat{c}^\dagger_{i\sigma} \hat{c}_{j\sigma} + U \sum_i \hat{n}_{i \uparrow} \hat{n}_{i\downarrow}, \\
\end{eqnarray}
where $\hat{c}^\dagger_{i\sigma}$ ($\hat{c}_{i\sigma}$) is the creation (annihilation) operator of a fermion with spin $\sigma\in \{ \uparrow,\downarrow\}$ at $i$-th site, and $\hat{n}_{i\sigma} =\hat{c}^\dagger_{i\sigma} \hat{c}_{i\sigma} $ is the particle number operator.
Parameters $\mu$, $U$, $h$ denotes the chemical potential, the on-site Coulomb interaction, and the external magnetic field, respectively.
The first term describes the kinetic part (cf. Ref.~\cite{cichy.kapcia.22}).
Here, we consider that hopping only between nearest neighbors (with $t_{ij} \equiv t >0$ as energy unit) and between NNN (with $t_{ij} \equiv t'$).
In the work, we restricts our investigation to the case of attractive $U<0$, which is a source of superconductivity in the system. 
The interaction term in such a case, after the mean-field decoupling, leads to the BCS-like term \cite{micnas.ranninger.90,RobaszkiewiczPRB1981A,RobaszkiewiczPRB1981B,ACichyEPL,ACichyAoP}:
\begin{eqnarray}\label{eq.MFAham}
    \hat{H}_{SC} = U \sum_{i} \left[ \left( \Delta_i \hat{c}^\dagger_{i\uparrow} \hat{c}^\dagger_{i\downarrow} + H.c. \right) - |\Delta_i| ^2 \right] ,
\end{eqnarray}
which describes $s$-wave superconductivity. 
The following order parameter $\Delta_i = \langle \hat{c}_{i\downarrow} \hat{c}_{i\uparrow} \rangle $ 
is a superconducting order parameter (here, the spatially homogeneous system is assumed, i.e., $\Delta_i \equiv \Delta$).
To find the ground state solutions we minimize the grand canonical potential defined as:
\begin{eqnarray}
    \Omega \equiv \Omega(\Delta) = -k_BT \ln \left\{\textrm{Tr} \left[ \exp \left( - \hat{H}/(k_BT)\right)\right] \right\}
\end{eqnarray}
 with respect to $\Delta$ and fixing other model parameters ($\mu$, $t'$, and $U$).
This procedure allows us to find order parameter $\Delta$, total number of particles $n$, and magnetization $m$ (a difference between number of spin-$\uparrow$ and spin-$\downarrow$ particles) from the following equations:
\begin{eqnarray}
    \frac{ \partial \Omega }{ \partial \Delta } = 0, \quad n = - \frac{ \partial \Omega }{ \partial \mu } , \quad \text{and} \quad m = - \frac{\partial \Omega }{ \partial h } , \label{eq.eqs}
\end{eqnarray}
respectively. 
More details can be found in Ref.~\cite{ptok.cichy.17}.

\begin{figure}[t]
    \centering
    \includegraphics[width=\wymiarsmall]{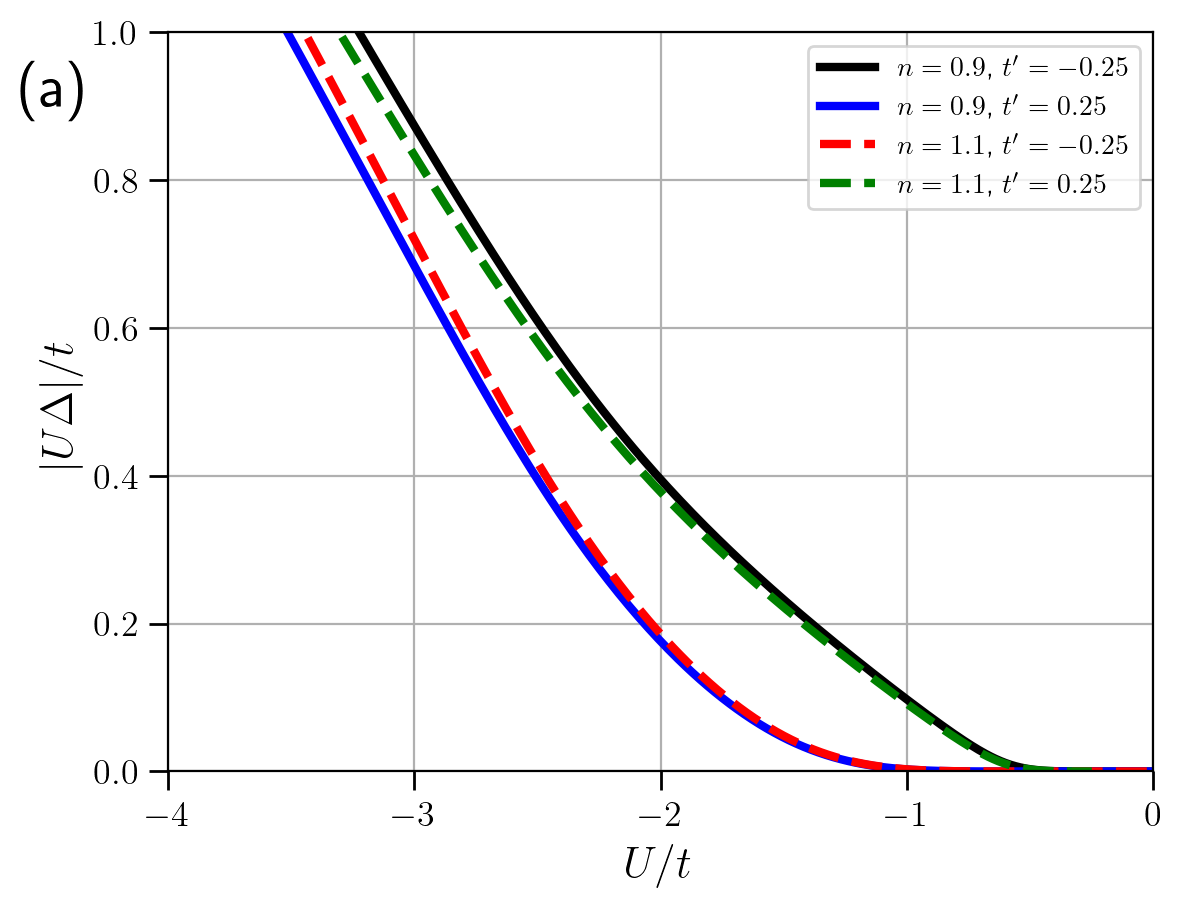}
    \includegraphics[width=\wymiarsmall]{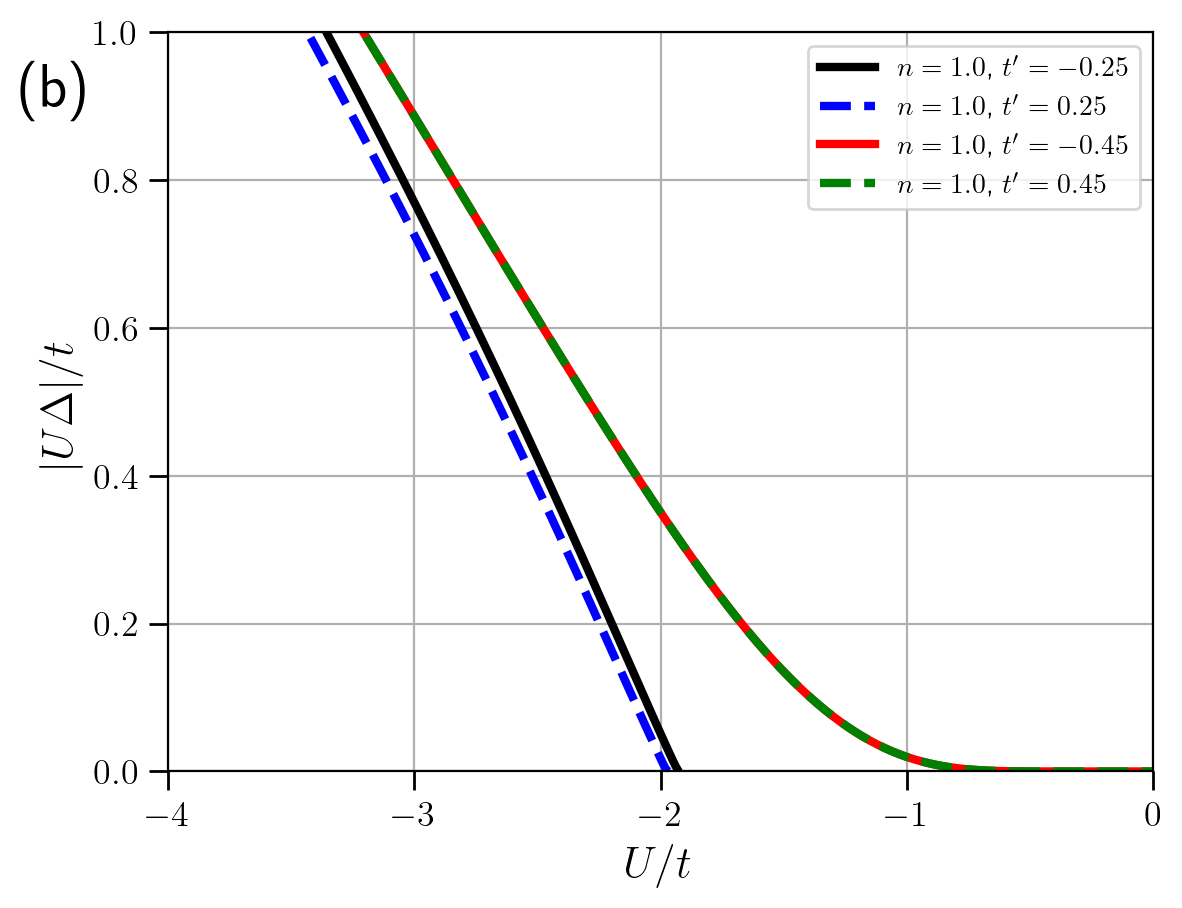}
    \includegraphics[width=\wymiarsmall]{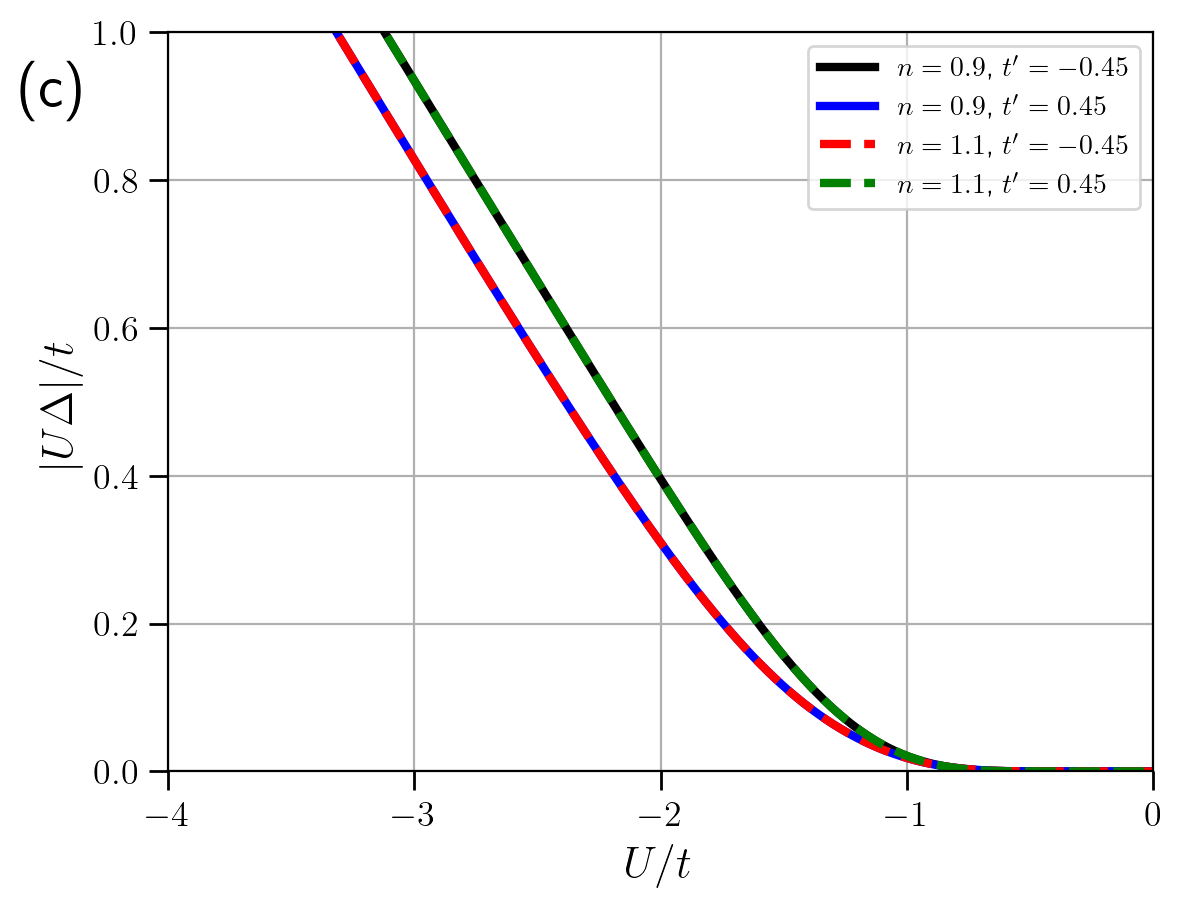}
    \caption{\label{fig.trans}%
        The dependence of the order parameter $|U\Delta|/t$ as a function of on-site interaction $U$ for various model parameters (as labeled) in the absence of the external field ($h=0$).}
\end{figure}

\section{Results}
\label{sec.res}

Let us start our investigation with the non-interacting electronic density of states (DOS), which is presented in Fig.~\ref{fig.dos}.
For convenient comparison of the DOSs for different $t'$, we present them in such a way, that the half-filling condition corresponds to $\omega = 0$.
The honeycomb lattice without NNN hopping ($t' = 0$) possess the DOS, which is symmetric with respect to the half-filling.
For $\omega = \pm t$, the Van Hove singularities (VHS) occur, while for $\omega^{\pm} \rightarrow 0$ the DOS disappear due to the Dirac cones existence.
For finite $t'$, the DOS lost its symmetric form, and the VHSs are splitted up.
The increase of the hopping between next nearest neighbours sites $t'$, initially leads to modification of the DOS around the ``oryginal'' VHS (for small, $|t'| < t/3$).
For $t'$ large enough (i.e., $| t' | > t/3$), modification of the DOS is well visible also around $\omega = 0$ (e.g. red line in Fig.~\ref{fig.dos}).
Transition from zero to finite value of DOS at $\omega=0$ occurs directly for $| t' | = t/3$.
What is interesting, the (non)zero DOS at $\omega = 0$ is related to the type of the phase transition from semiconducting to superconducting phase realized in the system.
In the next paragraph, we show that this value of $t'$ has crucial role on physical properties of the honeycomb lattice.

The type of the phase transition from semiconducting to the superconducting one at the ground state, can be clearly seen in the \mbox{$U$-dependence} superconducting order parameter $\Delta$ (Fig.~\ref{fig.trans}).
Let us start discussions from a case, when the external magnetic field is absent ($h=0$).
In the case of the half-filing ($n=1$, Fig.~\ref{fig.trans}(b)), a continuous phase transition from the superconducting (the BCS phase) to the normal state is related to the almost linear dependence of $\Delta$.
This type of behavior was earlier reported for the pure honeycomb lattice~\cite{zhao.paramekanti.06}, and preserved as long as $| t' | < t/3$. 
In fact, the existence of this phase transition is restricted to the cases where the DOS for $\omega=0$ disappears.
For such cases, the superconducting phase can be realized for pairing interaction stronger than some finite critical one $|U_{c}|$.
Indeed, for $| t' | > t/3$ or away from half-filling (Figs.~\ref{fig.trans}(a), and~\ref{fig.trans}(c)), the superconducting phase is stable for any $U<0$ (which corresponds to the exponential decay of $\Delta$ for $U \rightarrow 0$).
For $t'=0$ in the superconducting phase (with $\Delta \neq 0$), the magnetization is zero ($m=0$). 
Note also that presented curves in Fig.~\ref{fig.trans} clearly shows that the investigated model exhibits asymmetry with respect to the half-filling (results for $n<1$ and $n'=2-n>1$ are different) as well as for asymmetry with respect to change $t'$ into $-t'$.
However, the asymmetry $n \leftrightarrow n'=2-n$ is less pronounced for larger $t'$ (cf. Fig.~\ref{fig.trans}(c)).

\subsection*{The presence of the external magnetic field}

\begin{figure}[th!]
    \centering
    \includegraphics[width=\wymiarsmall]{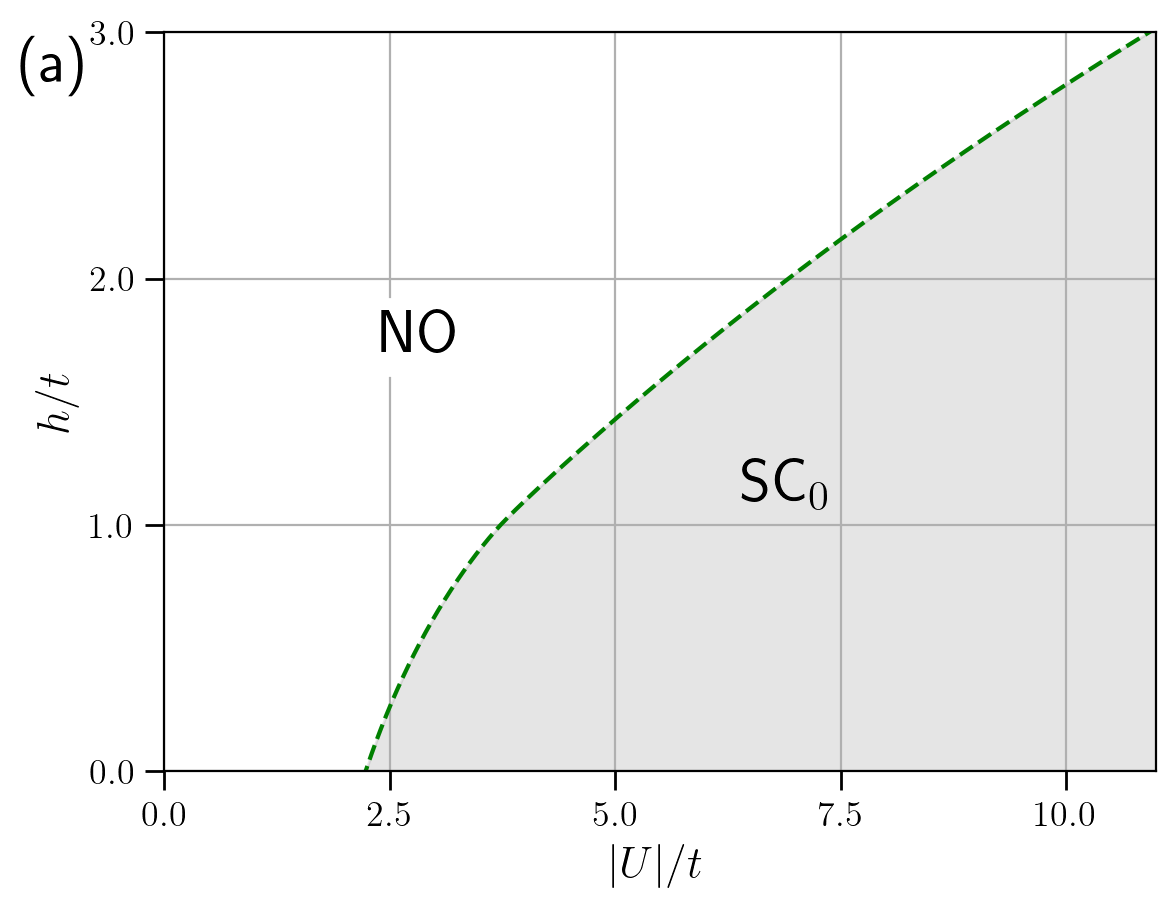}
    \includegraphics[width=\wymiarsmall]{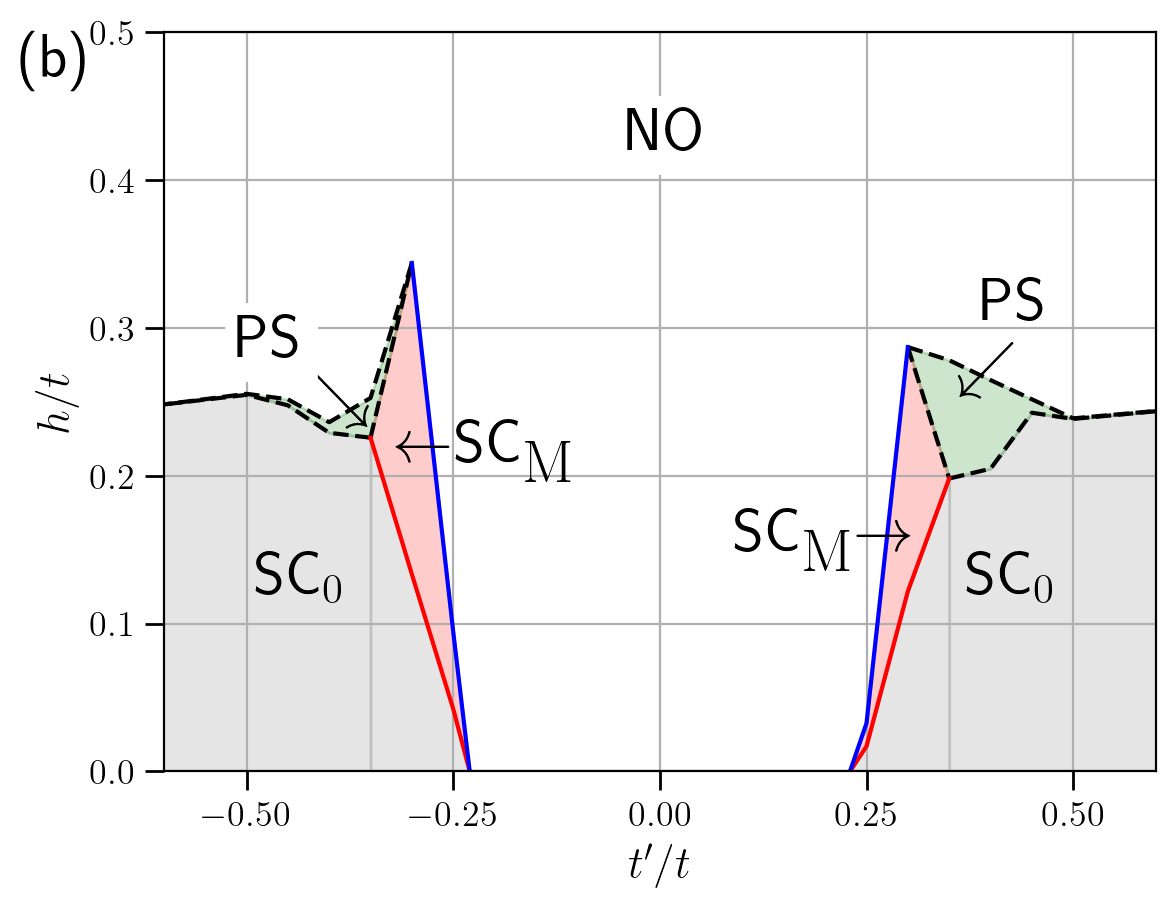}
    \caption{\label{fig.PDSARMA}%
        (a) The ground state phase diagram for $t'=0.0$ and $n=1$. 
        (b) The ground state phase diagram for $U/t=-2.0$ and $n=1$. 
        Symbols NO, SC$_0$, SC$_\textrm{M}$, and PS denote normal state, BCS-like phase, magnetized superconducting state and phase separation, respectively.}
\end{figure}

Now, we discuss the role of the magnetic field $h$ in described system.
In the presence of the magnetic field, the gap equation (the left expression in Eq.~(\ref{eq.eqs})) can have two nonequivalent solutions with $\Delta \neq 0$.
One of them, related to the (non-polarized magnetically) BCS-like phase, corresponds to $\Delta$, which is independent of $h$.
The other solution, related to phase called {\it Sarma phase} \cite{sarma.63,cichy.kapcia.22}, corresponds to the spin-polarized phase, for which $\Delta$ exhibit strong $h$ dependence.
For a typical situation, in the weak coupling region, the BCS phase has usually lower energy then than Sarma phase.
However, introduced $t' \neq 0$ can lead to stabilization of the spin-polarized superconducting state (in the calculations, only the Cooper pairs with zero total momentum are considered, i.e., we do not introduce the FFLO phase, where the Cooper pairs have non-zero total momentum,  e.g., Refs. ~\cite{ptok.cichy.17,cichy.ptok.18,PtokJSNM2018}).
Indeed, the results presented below clearly show that the Sarma phase can be stable in some range of model parameters.

The resulting phase diagram at the half-filing for $t'=0$ is presented in Fig.~\ref{fig.PDSARMA}(a).
The BSC-like superconducting phase (SC$_0$) is the only superconducting phase existing in the diagram.
It can exist for $|U|/t>|U_c|/t \approx 2.23$.
The value of magnetic field destroying superconductivity increases with $|U|/t$. 
The transition to the normal phase is a discontinuous one for $h\neq 0$, whereas it is continuous for $h = 0$.

In Fig.~\ref{fig.PDSARMA}(b), the phase diagram for $U/t=-2.0$ and $n=1$ is presented.
In the center of the diagram, for small $t'/t$, only the normal phase exists. 
With further increase of $|t'|/t$, both superconducting phases appear (the BCS phase for low $h/t$ and the Sarma phase for larger $h/t$).
For higher $|t'|/t$ the Sarma phase is destroyed and between the BCS phase and the NO phase the regions of (macroscopic) phase separation exist (coexistence of the NO phase and the BCS phase in two domains).

\begin{figure}[t]
    \centering
    \includegraphics[width=\wymiarsmall]{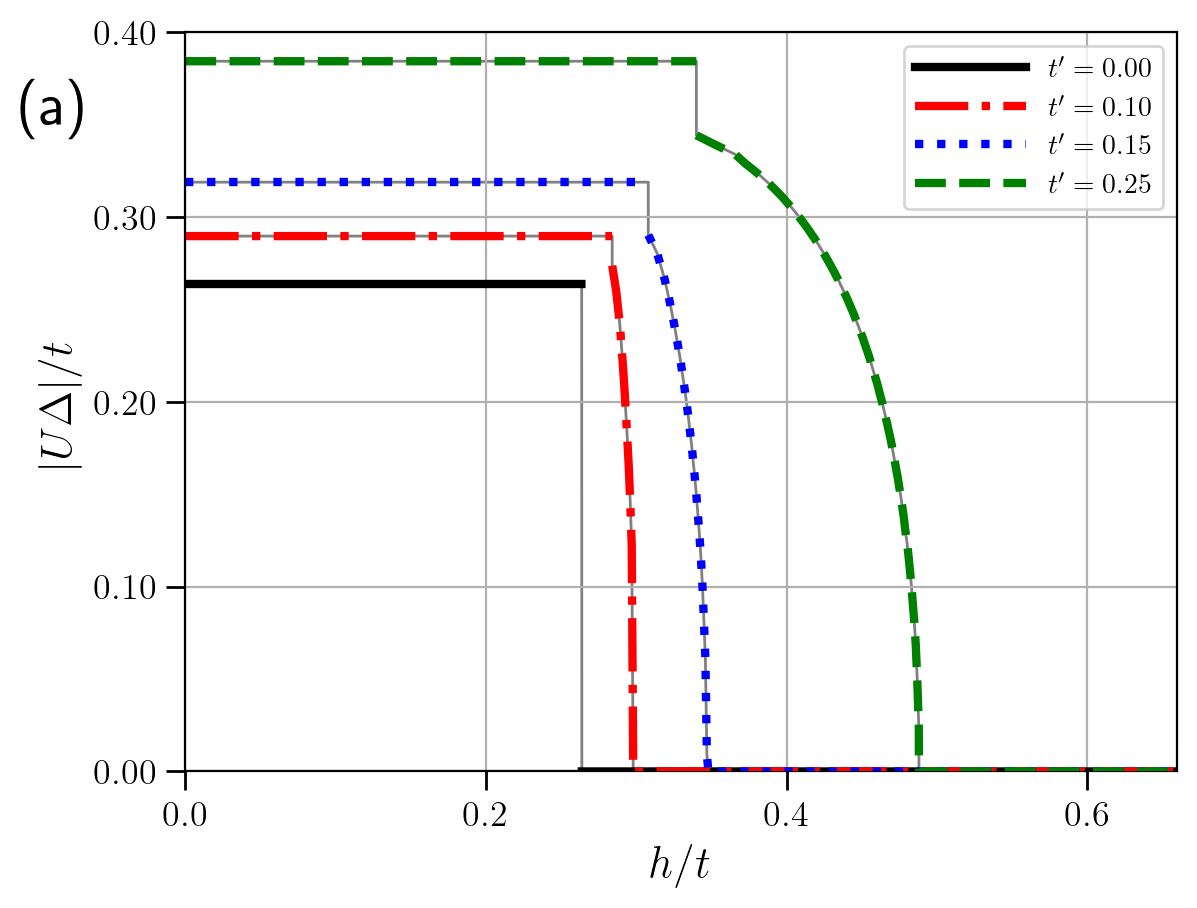}
    \includegraphics[width=\wymiarsmall]{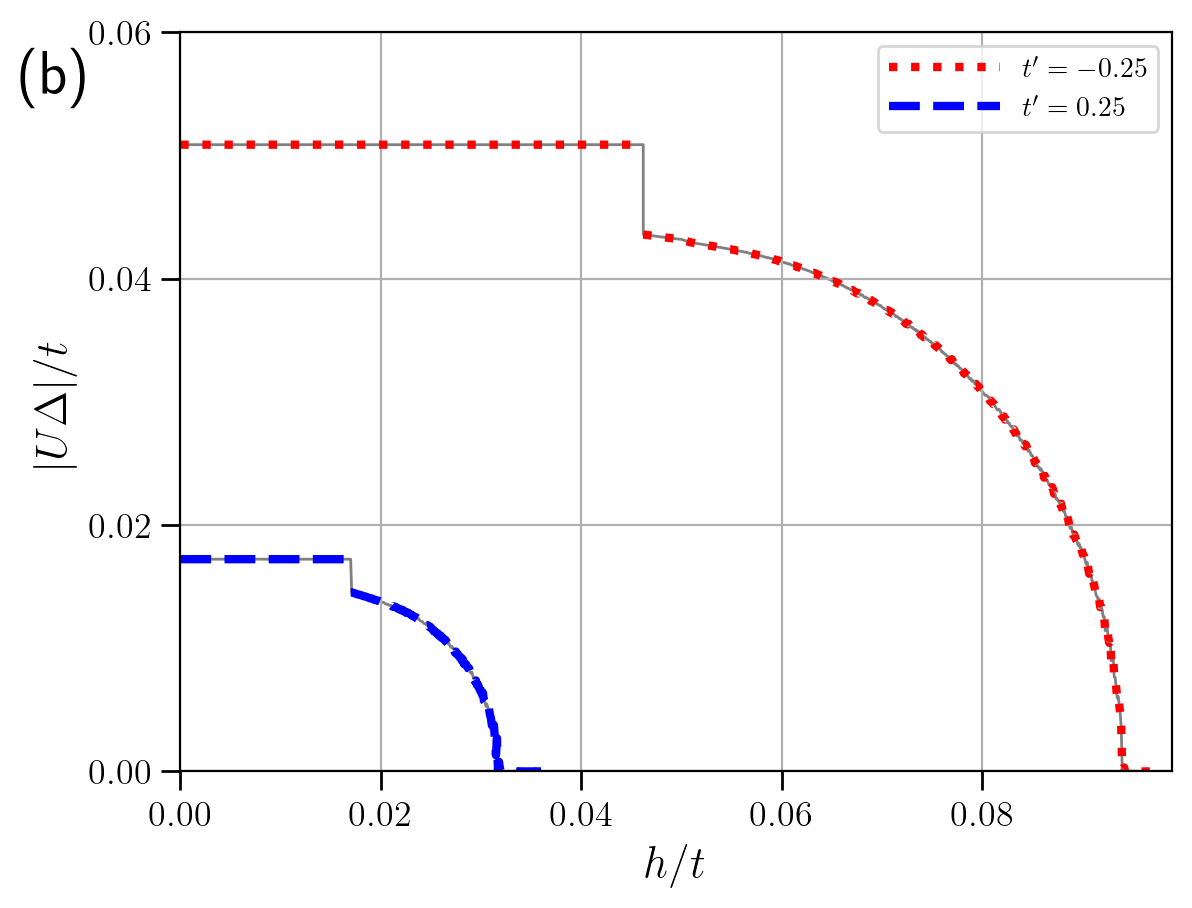}
    \caption{\label{fig.mag}%
        The dependence of order parameter $|U\Delta|/t$ as a function of external field $h$ at half-filling ($n=1$) and for (a) $U/t=-2.5$ and (b) $U/t=-2.0$ and various values of $t'$ (as labeled).}
\end{figure}

The exemplary curves for dependence of $\Delta$ as a function of $h$ at the ground state, are presented in Fig. \ref{fig.mag}.
It is clearly visible that increasing of the magnetic field destabilizes superconducting state (cf. also \cite{kapcia.robaszkiewicz.13,kapcia.14}).
Additionally, for $t' \neq 0$ and with increasing $h$, the discontinuous transition between two different superconducting phases occurs.
For low $h$ the phase with \mbox{$m = 0$} (i.e., the BCS phase) is stable with $\Delta$ independent of $h$, whereas above the transition point, the spin-polarized superconducting phase (with $m \neq 0$, the Sarma phase \cite{sarma.63,cichy.kapcia.22}) is stable, cf.~Fig.~\ref{fig.mag}. 
The transition between two superconducting phases occurring in non-zero field is discontinuous.
Further increasing of $h$ leads to a continuous transition from spin-polarized superconducting phase to the normal state.
One notices that increasing of $|t'|$ extends the regions of superconducting phases occurrence (Fig.~\ref{fig.mag}(a)).
From Fig.~\ref{fig.mag}(b), it is clearly visible that for $t'<0$ these regions are wider that for $t'$ of the opposite sign, which is also in agreement with the results from Fig.~\ref{fig.PDSARMA}(b) (at least for $U/t=-2$; for $U/t=-2.5$ differences are smaller,  cf. also Fig.~3 from Ref.~\cite{cichy.kapcia.22}).

\begin{figure}[th!]
    \centering
    \includegraphics[width=\wymiarsmall]{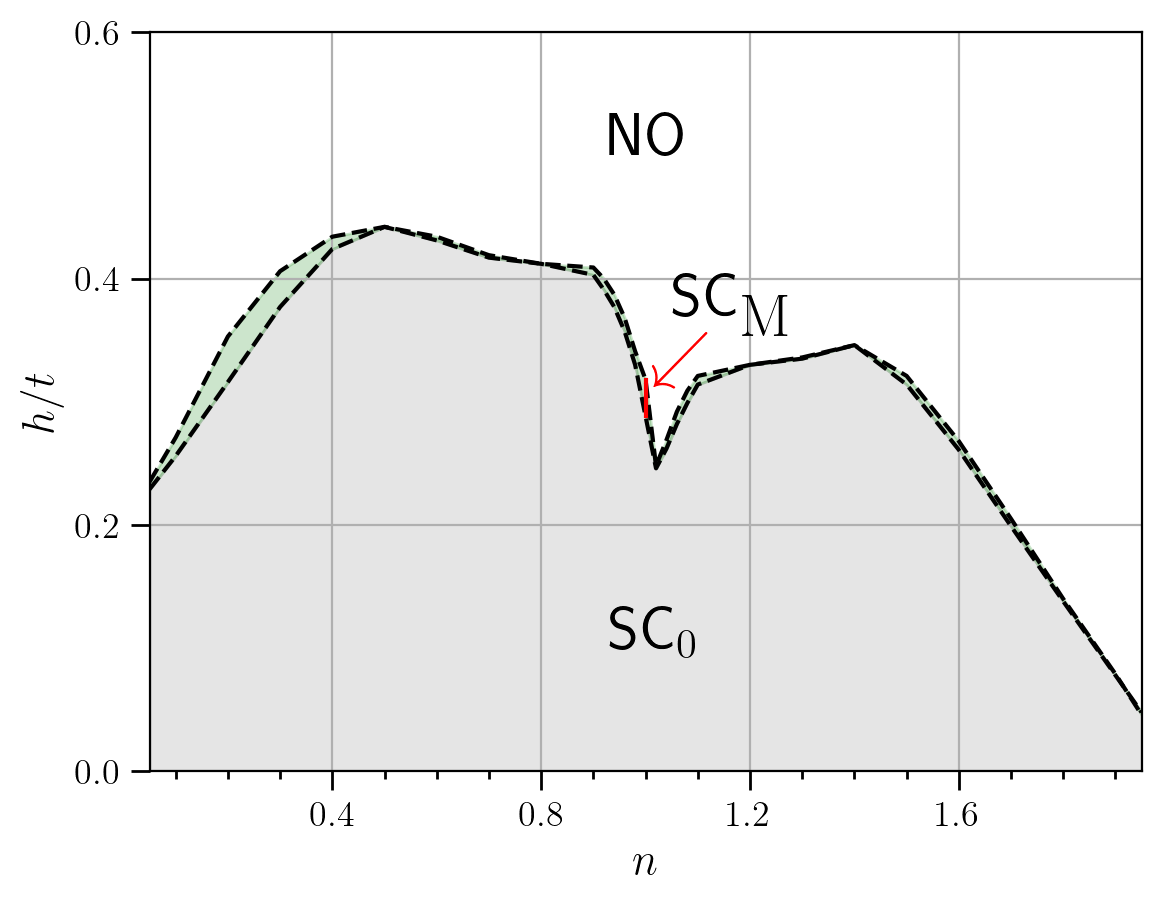}
    \caption{\label{fig.diagramfunkn}%
        The phase diagram for $U/t=-2.5$ and $t'/t=-0.1$. Symbols as in Fig. \ref{fig.PDSARMA}.}
\end{figure}

To have overall picture of the behavior of studied model as a function of the total concentration $n$, the $h/t$ vs. $n$ phase diagram for $t'/t = -0.1$ and $U/t=-2.5$ is presented in Fig.~\ref{fig.diagramfunkn}. 
As mentioned previously, the Sarma phase occurs in a very narrow region only at $n=1$ for $h \neq 0$. 
In the absence of magnetic field, for these model parameters, the ground state is the BSC-like superconductor.
The magnetic field destroys the BCS-like phase, however, the critical field depends on $n$.

\section{Summary}
\label{sec.sum}

In this work, we investigated the role of the hopping between next-nearest-neighbor sites on the type of phase transition between semiconducting and superconducting phases at close vicinity of the half-filling in the honeycomb lattice (at the ground state).
First, we showed that the occurrence of the phase transition from semiconducting (semimetallic) to superconducting states depends on the next nearest-neighbor hopping.
For the hopping integral larger than third part of nearest-neighbor hopping, the phase transition between the mentioned phases does not occur (the superconducting phase is stable for any $U$), whereas for $| t' | < t/3$ some critical value of $U$ exists and the transition is a continuous one. 
Additionally, we investigated the dependence of the superconducting order parameter $\Delta$ as a function of the external magnetic field. 
Independently of the hopping between next-nearest-neighbor sites, we observed the discontinuous phase transition from the BCS superconducting phase (with $\Delta$ independent of magnetic field) to the Sarma phase (with $\Delta$ dependent on magnetic field).
The range of magnetic field, for which the Sarma phase exists, strongly depends on the system parameters.
However, the Sarma phase cannot occur for $|t'|>t/3$ and $n\neq 1$ (away from the half-filling).
The occurrence of the Sarma phase is associated to the semiconducting features of non-interacting band structure for the honeycomb lattice, particularly with the existence of the Dirac cones at the Fermi level for half-filling.

\section*{Acknowledgements}
We kindly thank the late Professor Roman Micnas fol all his inspiration and motivation that he provided for this work.
This work was supported by National Science Centre (NCN, Poland) under Project
Nos. UMO-2017/24/C/ST3/00357 (A.C.), and UMO-2021/43/B/ST3/02166 (A.P.). 
K.J.K. thanks the Polish National Agency for Academic Exchange for funding in the frame of the Bekker programme (PPN/BEK/2020/1/00184).
A.P. is grateful to Laboratoire de Physique des Solides in Orsay (CNRS, University Paris Saclay) for hospitality during a part of the
work on this project.
Access to computing and storage facilities provided by the Pozna\'n Supercomputing and Networking Center (EAGLE cluster) is greatly appreciated.

\section*{Declaration of Competing Interest}
The authors declares that they have no known competing financial interests or personal relationships that could have appeared to influence the work reported in this paper.
The funders had no role in the design of the study; in the collection, analyses, or interpretation of data; in the writing of the manuscript, or in the decision to publish the results.

\printcredits

\bibliographystyle{elsarticle-num}

\bibliography{bibliography}

\begin{thebibliography}{10}
\expandafter\ifx\csname url\endcsname\relax
  \def\url#1{\texttt{#1}}\fi
\expandafter\ifx\csname urlprefix\endcsname\relax\def\urlprefix{URL }\fi
\expandafter\ifx\csname href\endcsname\relax
  \def\href#1#2{#2} \def\path#1{#1}\fi

\bibitem{castroneto.guinea.09}
A.~H. Castro~Neto, F.~Guinea, N.~M.~R. Peres, K.~S. Novoselov, A.~K. Geim,
  \href{https://doi.org/10.1103/RevModPhys.81.109}{The electronic properties of
  graphene}, Rev. Mod. Phys. 81~(1) (2009) 109--162.
\newblock \href {https://doi.org/10.1103/RevModPhys.81.109}
  {\path{doi:10.1103/RevModPhys.81.109}}.
\newline\urlprefix\url{https://doi.org/10.1103/RevModPhys.81.109}

\bibitem{ryu.hatsugai.02}
S.~Ryu, Y.~Hatsugai,
  \href{https://doi.org/10.1103/PhysRevLett.89.077002}{Topological origin of
  zero-energy edge states in particle-hole symmetric systems}, Phys. Rev. Lett.
  89~(7) (2002) 077002.
\newblock \href {https://doi.org/10.1103/PhysRevLett.89.077002}
  {\path{doi:10.1103/PhysRevLett.89.077002}}.
\newline\urlprefix\url{https://doi.org/10.1103/PhysRevLett.89.077002}

\bibitem{wakabayashi.takane.07}
K.~Wakabayashi, Y.~Takane, M.~Sigrist,
  \href{https://doi.org/10.1103/PhysRevLett.99.036601}{Perfectly conducting
  channel and universality crossover in disordered graphene nanoribbons}, Phys.
  Rev. Lett. 99~(3) (2007) 036601.
\newblock \href {https://doi.org/10.1103/PhysRevLett.99.036601}
  {\path{doi:10.1103/PhysRevLett.99.036601}}.
\newline\urlprefix\url{https://doi.org/10.1103/PhysRevLett.99.036601}

\bibitem{yao.yang.09}
W.~Yao, S.~A. Yang, Q.~Niu,
  \href{https://doi.org/10.1103/PhysRevLett.102.096801}{Edge states in
  graphene: From gapped flat-band to gapless chiral modes}, Phys. Rev. Lett.
  102~(9) (2009) 096801.
\newblock \href {https://doi.org/10.1103/PhysRevLett.102.096801}
  {\path{doi:10.1103/PhysRevLett.102.096801}}.
\newline\urlprefix\url{https://doi.org/10.1103/PhysRevLett.102.096801}

\bibitem{kane.mele.05}
C.~L. Kane, E.~J. Mele,
  \href{https://doi.org/10.1103/PhysRevLett.95.146802}{${Z}_{2}$ topological
  order and the quantum spin {Hall} effect}, Phys. Rev. Lett. 95~(14) (2005)
  146802.
\newblock \href {https://doi.org/10.1103/PhysRevLett.95.146802}
  {\path{doi:10.1103/PhysRevLett.95.146802}}.
\newline\urlprefix\url{https://doi.org/10.1103/PhysRevLett.95.146802}

\bibitem{zhang.tan.05}
Y.~Zhang, Y.-W. Tan, H.~L. Stormer, P.~Kim,
  \href{https://doi.org/10.1038/nature04235}{Experimental observation of the
  quantum {Hall} effect and {Berry's} phase in graphene}, Nature 438 (2005)
  201--204.
\newblock \href {https://doi.org/10.1038/nature04235}
  {\path{doi:10.1038/nature04235}}.
\newline\urlprefix\url{https://doi.org/10.1038/nature04235}

\bibitem{avsar.ochoa.20}
A.~Avsar, H.~Ochoa, F.~Guinea, B.~\"Ozyilmaz, B.~J. van Wees, I.~J. Vera-Marun,
  \href{https://doi.org/10.1103/RevModPhys.92.021003}{Colloquium: Spintronics
  in graphene and other two-dimensional materials}, Rev. Mod. Phys. 92~(2)
  (2020) 021003.
\newblock \href {https://doi.org/10.1103/RevModPhys.92.021003}
  {\path{doi:10.1103/RevModPhys.92.021003}}.
\newline\urlprefix\url{https://doi.org/10.1103/RevModPhys.92.021003}

\bibitem{bussolotti.kawai.18}
F.~Bussolotti, H.~Kawai, Z.~E. Ooi, V.~Chellappan, D.~Thian, A.~L.~C. Pang,
  K.~E.~J. Goh, \href{https://doi.org/10.1088/2399-1984/aac9d7}{Roadmap on
  finding chiral valleys: screening {2D} materials for valleytronics}, Nano
  Futures 2~(3) (2018) 032001.
\newblock \href {https://doi.org/10.1088/2399-1984/aac9d7}
  {\path{doi:10.1088/2399-1984/aac9d7}}.
\newline\urlprefix\url{https://doi.org/10.1088/2399-1984/aac9d7}

\bibitem{bistritzer.macdonald.11}
R.~Bistritzer, A.~H. MacDonald,
  \href{https://doi.org/10.1073/pnas.1108174108}{Moir{\'e} bands in twisted
  double-layer graphene}, PNAS 108~(30) (2011) 12233--12237.
\newblock \href {https://doi.org/10.1073/pnas.1108174108}
  {\path{doi:10.1073/pnas.1108174108}}.
\newline\urlprefix\url{https://doi.org/10.1073/pnas.1108174108}

\bibitem{he.zhou.21}
F.~He, Y.~Zhou, Z.~Ye, S.-H. Cho, J.~Jeong, X.~Meng, Y.~Wang,
  \href{https://doi.org/10.1021/acsnano.0c10435}{Moir{\'e} patterns in {2D}
  materials: A review}, ACS Nano 15~(4) (2021) 5944--5958.
\newblock \href {https://doi.org/10.1021/acsnano.0c10435}
  {\path{doi:10.1021/acsnano.0c10435}}.
\newline\urlprefix\url{https://doi.org/10.1021/acsnano.0c10435}

\bibitem{cao.fatemi.18}
Y.~Cao, V.~Fatemi, S.~Fang, K.~Watanabe, T.~Taniguchi, E.~Kaxiras,
  P.~Jarillo-Herrero, \href{https://doi.org/10.1038/nature26160}{Unconventional
  superconductivity in magic-angle graphene superlattices}, Nature 556 (2018)
  43--50.
\newblock \href {https://doi.org/10.1038/nature26160}
  {\path{doi:10.1038/nature26160}}.
\newline\urlprefix\url{https://doi.org/10.1038/nature26160}

\bibitem{lu.setpanov.19}
X.~Lu, P.~Stepanov, W.~Yang, M.~Xie, M.~A. Aamir, I.~Das, C.~Urgell,
  K.~Watanabe, T.~Taniguchi, G.~Zhang, A.~Bachtold, A.~H. MacDonald, D.~K.
  Efetov, \href{https://doi.org/10.1038/s41586-019-1695-0}{Superconductors,
  orbital magnets and correlated states in magic-angle bilayer graphene},
  Nature 574 (2019) 653--657.
\newblock \href {https://doi.org/10.1038/s41586-019-1695-0}
  {\path{doi:10.1038/s41586-019-1695-0}}.
\newline\urlprefix\url{https://doi.org/10.1038/s41586-019-1695-0}

\bibitem{yankowitz.shaowen.19}
M.~Yankowitz, S.~Chen, H.~Polshyn, Y.~Zhang, K.~Watanabe, T.~Taniguchi,
  D.~Graf, A.~F. Young, C.~R. Dean,
  \href{https://doi.org/10.1126/science.aav1910}{Tuning superconductivity in
  twisted bilayer graphene}, Science 363~(6431) (2019) 1059--1064.
\newblock \href {https://doi.org/10.1126/science.aav1910}
  {\path{doi:10.1126/science.aav1910}}.
\newline\urlprefix\url{https://doi.org/10.1126/science.aav1910}

\bibitem{arora.polski.20}
H.~S. Arora, R.~Polski, Y.~Zhang, A.~Thomson, Y.~Choi, H.~Kim, Z.~Lin, I.~Z.
  Wilson, X.~Xu, J.-H. Chu, K.~Watanabe, T.~Taniguchi, J.~Alicea,
  S.~Nadj-Perge,
  \href{https://doi.org/10.1038/s41586-020-2473-8}{Superconductivity in
  metallic twisted bilayer graphene stabilized by {WSe$_{2}$}}, Nature 583
  (2020) 379--384.
\newblock \href {https://doi.org/10.1038/s41586-020-2473-8}
  {\path{doi:10.1038/s41586-020-2473-8}}.
\newline\urlprefix\url{https://doi.org/10.1038/s41586-020-2473-8}

\bibitem{codecido.wang.19}
E.~Codecido, Q.~Wang, R.~Koester, S.~Che, H.~Tian, R.~Lv, S.~Tran, K.~Watanabe,
  T.~Taniguchi, F.~Zhang, M.~Bockrath, C.~N. Lau,
  \href{https://doi.org/10.1126/sciadv.aaw9770}{Correlated insulating and
  superconducting states in twisted bilayer graphene below the magic angle},
  Sci. Adv. 5 (2019) eaaw9770.
\newblock \href {https://doi.org/10.1126/sciadv.aaw9770}
  {\path{doi:10.1126/sciadv.aaw9770}}.
\newline\urlprefix\url{https://doi.org/10.1126/sciadv.aaw9770}

\bibitem{saito.ge.20}
Y.~Saito, J.~Ge, K.~Watanabe, T.~Taniguchi, A.~F. Young,
  \href{https://doi.org/10.1038/s41567-020-0928-3}{Independent superconductors
  and correlated insulators in twisted bilayer graphene}, Nat. Phys. 16 (2020)
  926--930.
\newblock \href {https://doi.org/10.1038/s41567-020-0928-3}
  {\path{doi:10.1038/s41567-020-0928-3}}.
\newline\urlprefix\url{https://doi.org/10.1038/s41567-020-0928-3}

\bibitem{stepanov.das.20}
P.~Stepanov, I.~Das, X.~Lu, A.~Fahimniya, K.~Watanabe, T.~Taniguchi, F.~H.~L.
  Koppens, J.~Lischner, L.~Levitov, D.~K. Efetov,
  \href{https://doi.org/10.1038/s41586-020-2459-6}{Untying the insulating and
  superconducting orders in magic-angle graphene}, Nature 583 (2020) 375--378.
\newblock \href {https://doi.org/10.1038/s41586-020-2459-6}
  {\path{doi:10.1038/s41586-020-2459-6}}.
\newline\urlprefix\url{https://doi.org/10.1038/s41586-020-2459-6}

\bibitem{cao.fatemi.18b}
Y.~Cao, V.~Fatemi, A.~Demir, S.~Fang, S.~L. Tomarken, J.~Y. Luo, J.~D.
  Sanchez-Yamagishi, K.~Watanabe, T.~Taniguchi, E.~Kaxiras, R.~C. Ashoori,
  P.~Jarillo-Herrero, \href{https://doi.org/10.1038/nature26154}{Correlated
  insulator behaviour at half-filling in magic-angle graphene superlattices},
  Nature 556 (2018) 80--84.
\newblock \href {https://doi.org/10.1038/nature26154}
  {\path{doi:10.1038/nature26154}}.
\newline\urlprefix\url{https://doi.org/10.1038/nature26154}

\bibitem{zhang.macdonald.13}
F.~Zhang, A.~H. MacDonald, E.~J. Mele,
  \href{https://www.pnas.org/content/110/26/10546}{Valley {Cher}n numbers and
  boundary modes in gapped bilayer graphene}, PNAS 110 (2013) 10546.
\newblock \href {https://doi.org/10.1073/pnas.1308853110}
  {\path{doi:10.1073/pnas.1308853110}}.
\newline\urlprefix\url{https://www.pnas.org/content/110/26/10546}

\bibitem{vaezi.liang.13}
A.~Vaezi, Y.~Liang, D.~H. Ngai, L.~Yang, E.-A. Kim,
  \href{https://doi.org/10.1103/PhysRevX.3.021018}{Topological edge states at a
  tilt boundary in gated multilayer graphene}, Phys. Rev. X 3~(2) (2013)
  021018.
\newblock \href {https://doi.org/10.1103/PhysRevX.3.021018}
  {\path{doi:10.1103/PhysRevX.3.021018}}.
\newline\urlprefix\url{https://doi.org/10.1103/PhysRevX.3.021018}

\bibitem{ju.shi.15}
L.~Ju, Z.~Shi, N.~Nair, Y.~Lv, C.~Jin, J.~Velasco, C.~Ojeda-Aristizabal, H.~A.
  Bechtel, M.~C. Martin, A.~Zettl, J.~Analytis, F.~Wang,
  \href{https://doi.org/10.1038/nature14364}{Topological valley transport at
  bilayer graphene domain walls}, Nature 520 (2015) 650--655.
\newblock \href {https://doi.org/10.1038/nature14364}
  {\path{doi:10.1038/nature14364}}.
\newline\urlprefix\url{https://doi.org/10.1038/nature14364}

\bibitem{brown.walet.18}
R.~Brown, N.~R. Walet, F.~Guinea,
  \href{https://doi.org/10.1103/PhysRevLett.120.026802}{Edge modes and nonlocal
  conductance in graphene superlattices}, Phys. Rev. Lett. 120~(2) (2018)
  026802.
\newblock \href {https://doi.org/10.1103/PhysRevLett.120.026802}
  {\path{doi:10.1103/PhysRevLett.120.026802}}.
\newline\urlprefix\url{https://doi.org/10.1103/PhysRevLett.120.026802}

\bibitem{hunt.sanchezyamagishi.13}
B.~Hunt, J.~D. Sanchez-Yamagishi, A.~F. Young, M.~Yankowitz, B.~J. LeRoy,
  K.~Watanabe, T.~Taniguchi, P.~Moon, M.~Koshino, P.~Jarillo-Herrero, R.~C.
  Ashoori, \href{https://doi.org/10.1126/science.1237240}{Massive {Dirac}
  fermions and {Hofstadter} butterfly in a van der {Waals} heterostructure},
  Science 340~(6139) (2013) 1427--1430.
\newblock \href {https://doi.org/10.1126/science.1237240}
  {\path{doi:10.1126/science.1237240}}.
\newline\urlprefix\url{https://doi.org/10.1126/science.1237240}

\bibitem{zhao.paramekanti.06}
E.~Zhao, A.~Paramekanti,
  \href{https://doi.org/10.1103/PhysRevLett.97.230404}{{BCS-BEC} crossover on
  the two-dimensional honeycomb lattice}, Phys. Rev. Lett. 97~(23) (2006)
  230404.
\newblock \href {https://doi.org/10.1103/PhysRevLett.97.230404}
  {\path{doi:10.1103/PhysRevLett.97.230404}}.
\newline\urlprefix\url{https://doi.org/10.1103/PhysRevLett.97.230404}

\bibitem{cichy.ptok.18}
A.~Cichy, A.~Ptok, \href{https://doi.org/10.1103/PhysRevA.97.053619}{Reentrant
  {Fulde--Ferrell--Larkin--Ovchinnikov} superfluidity in the honeycomb
  lattice}, Phys. Rev. A 97~(5) (2018) 053619.
\newblock \href {https://doi.org/10.1103/PhysRevA.97.053619}
  {\path{doi:10.1103/PhysRevA.97.053619}}.
\newline\urlprefix\url{https://doi.org/10.1103/PhysRevA.97.053619}

\bibitem{cichy.kapcia.22}
A.~Cichy, K.~J. Kapcia, A.~Ptok,
  \href{https://doi.org/10.1103/PhysRevB.105.214510}{Connection between the
  semiconductor-superconductor transition and the spin-polarized
  superconducting phase in the honeycomb lattice}, Phys. Rev. B 105~(21) (2022)
  214510.
\newblock \href {https://doi.org/10.1103/PhysRevB.105.214510}
  {\path{doi:10.1103/PhysRevB.105.214510}}.
\newline\urlprefix\url{https://doi.org/10.1103/PhysRevB.105.214510}

\bibitem{micnas.ranninger.90}
R.~Micnas, J.~Ranninger, S.~Robaszkiewicz,
  \href{https://doi.org/10.1103/RevModPhys.62.113}{Superconductivity in
  narrow-band systems with local nonretarded attractive interactions}, Rev.
  Mod. Phys. 62~(1) (1990) 113--171.
\newblock \href {https://doi.org/10.1103/RevModPhys.62.113}
  {\path{doi:10.1103/RevModPhys.62.113}}.
\newline\urlprefix\url{https://doi.org/10.1103/RevModPhys.62.113}

\bibitem{RobaszkiewiczPRB1981A}
S.~Robaszkiewicz, R.~Micnas, K.~A. Chao,
  \href{https://doi.org/10.1103/PhysRevB.24.4018}{Hartree theory for the
  negative-{$U$} extended {H}ubbard model: {G}round state}, Phys. Rev. B 24~(7)
  (1981) 4018--4024.
\newblock \href {https://doi.org/10.1103/PhysRevB.24.4018}
  {\path{doi:10.1103/PhysRevB.24.4018}}.
\newline\urlprefix\url{https://doi.org/10.1103/PhysRevB.24.4018}

\bibitem{RobaszkiewiczPRB1981B}
S.~Robaszkiewicz, R.~Micnas, K.~A. Chao,
  \href{https://doi.org/10.1103/PhysRevB.26.3915}{Hartree theory for the
  negative-{$U$} extended {H}ubbard model. {II.} {F}inite temperature}, Phys.
  Rev. B 26~(7) (1982) 3915--3922.
\newblock \href {https://doi.org/10.1103/PhysRevB.26.3915}
  {\path{doi:10.1103/PhysRevB.26.3915}}.
\newline\urlprefix\url{https://doi.org/10.1103/PhysRevB.26.3915}

\bibitem{ACichyEPL}
A.~Kujawa-Cichy, R.~Micnas,
  \href{https://doi.org/10.1209/0295-5075/95/37003}{Stability of superfluid
  phases in the {2D} spin-polarized attractive {H}ubbard model}, Europhys.
  Lett. 95~(3) (2011) 37003.
\newblock \href {https://doi.org/10.1209/0295-5075/95/37003}
  {\path{doi:10.1209/0295-5075/95/37003}}.
\newline\urlprefix\url{https://doi.org/10.1209/0295-5075/95/37003}

\bibitem{ACichyAoP}
A.~Cichy, R.~Micnas, \href{https://doi.org/10.1016/j.aop.2014.04.014}{The
  spin-imbalanced attractive {H}ubbard model in $d=3$: {P}hase diagrams and
  {BCS}–{BEC} crossover at low filling}, Ann. Physics 347 (2014) 207--249.
\newblock \href {https://doi.org/10.1016/j.aop.2014.04.014}
  {\path{doi:10.1016/j.aop.2014.04.014}}.
\newline\urlprefix\url{https://doi.org/10.1016/j.aop.2014.04.014}

\bibitem{ptok.cichy.17}
A.~Ptok, A.~Cichy, K.~Rodr\'{\i}guez, K.~J. Kapcia,
  \href{https://doi.org/10.1103/PhysRevA.95.033613}{Critical behavior in one
  dimension: Unconventional pairing, phase separation, {BEC-BCS} crossover, and
  magnetic {Lifshitz} transition}, Phys. Rev. A 95~(3) (2017) 033613.
\newblock \href {https://doi.org/10.1103/PhysRevA.95.033613}
  {\path{doi:10.1103/PhysRevA.95.033613}}.
\newline\urlprefix\url{https://doi.org/10.1103/PhysRevA.95.033613}

\bibitem{sarma.63}
G.~Sarma, \href{https://doi.org/10.1016/0022-3697(63)90007-6}{On the influence
  of a uniform exchange field acting on the spins of the conduction electrons
  in a superconductor}, J. Phys. Chem. Solids 24~(8) (1963) 1029--1032.
\newblock \href {https://doi.org/10.1016/0022-3697(63)90007-6}
  {\path{doi:10.1016/0022-3697(63)90007-6}}.
\newline\urlprefix\url{https://doi.org/10.1016/0022-3697(63)90007-6}

\bibitem{PtokJSNM2018}
A.~Ptok, A.~Cichy, K.~Rodr{\'\i}guez, K.~J. Kapcia,
  \href{https://doi.org/10.1007/s10948-017-4366-0}{Phase transitions in
  quasi-one-dimensional system with unconventional superconductivity}, J.
  Supercond. Nov. Magn. 31~(3) (2018) 697--702.
\newblock \href {https://doi.org/10.1007/s10948-017-4366-0}
  {\path{doi:10.1007/s10948-017-4366-0}}.
\newline\urlprefix\url{https://doi.org/10.1007/s10948-017-4366-0}

\bibitem{kapcia.robaszkiewicz.13}
K.~Kapcia, S.~Robaszkiewicz,
  \href{https://dx.doi.org/10.1088/0953-8984/25/6/065603}{The magnetic field
  induced phase separation in a model of a superconductor with local electron
  pairing}, J. Phys.: Condens. Matter 25~(6) (2013) 065603.
\newblock \href {https://doi.org/10.1088/0953-8984/25/6/065603}
  {\path{doi:10.1088/0953-8984/25/6/065603}}.
\newline\urlprefix\url{https://dx.doi.org/10.1088/0953-8984/25/6/065603}

\bibitem{kapcia.14}
K.~Kapcia,
  \href{http://dx.doi.org/10.12693/APhysPolA.126.A-53}{Superconductivity,
  metastability and magnetic field induced phase separation in the atomic limit
  of the {P}enson-{K}olb-{H}ubbard model}, Acta Phys. Pol. A 126~(4A) (2014)
  A--53--A--57.
\newblock \href {https://doi.org/10.12693/APhysPolA.126.A-53}
  {\path{doi:10.12693/APhysPolA.126.A-53}}.
\newline\urlprefix\url{http://dx.doi.org/10.12693/APhysPolA.126.A-53}

\end{thebibliography}
\end{document}